\documentclass[a4paper,11pt]{article}
\usepackage{arxiv}
\usepackage{geometry}

\usepackage[utf8]{inputenc} 
\usepackage[T1]{fontenc}    
\usepackage{hyperref}       
\usepackage{url}            
\usepackage{booktabs}       
\usepackage{amsfonts}       
\usepackage{nicefrac}       
\usepackage{microtype}      
\usepackage{lipsum}
\usepackage{graphicx}
\usepackage[numbers,sort&compress]{natbib}
\graphicspath{ {./images/} }
\usepackage{color}

\usepackage{authblk}

\title{\boldmath Evaluation of the KLauS ASIC at low temperature}

\author[a]{Wei Wang}
\author[b]{Wei Shen}
\author[b]{Zhenxiong Yuan}
\author[b]{Konrad Briggl}
\author[b]{Hans-Christian Schultz-Coulon}
\author[b]{Erik Warttmann}
\author[ \enspace c,f]{Wenqi Yan \thanks{Corresponding author(yanwq@ihep.ac.cn)}} 
\author[ \enspace c,d,f]{Guofu Cao \thanks{Corresponding author (caogf@ihep.ac.cn)}}
\author[e]{Zepeng Li}
\author[a]{Ming Qi}
\author[c,f]{Liangjian Wen}

\affil[a]{Nanjing University, Nanjing, 210093, China.}
\affil[b]{Kirchhoff Institut für Physik, Universität Heidelberg, Im Neuenheimer Feld 227, 69120, Germany.}
\affil[c]{Institute of High Energy Physics, Beijing, 100049, China.}
\affil[d]{University of Chinese Academy of Sciences, Beijing, China.}
\affil[f]{State Key Laboratory of Particle Detection and Electronics, Beijing 100049, China.}
\affil[e]{Wright Laboratory, Department of Physics, Yale University, New Haven, CT 06511, USA. }

\begin{document}

\maketitle
 
\begin{abstract}

The Taishan Antineutrino Observatory (TAO) is proposed to first use a cold liquid scintillator detector (-50~$^\circ$C) equipped with large-area silicon photomultipliers (SiPMs) ($\sim$10~m$^2$) to precisely measure the reactor antineutrino spectrum with a record energy resolution of < 2\% at 1~MeV. The KLauS ASIC shows excellent performance at room temperature and is a potential readout solution for TAO. In this work, we report evaluations of the fifth version of the KLauS ASIC (KLauS5) from room temperature to -50~$^\circ$C with inputs of injected charge or SiPMs. Our results show that KLauS5 has good performance at the tested temperatures with no significant degradation of the charge noise, charge linearity, gain uniformity or recovery time. Meanwhile, we also observe that several key parameters degrade when the chip operates in cold conditions, including the dynamic range and power consumption. However, even with this degradation, a good signal-to-noise ratio and good resolution of a single photoelectron can still be achieved for the tested SiPM with a gain of greater than 1.5$\times$10$^6$ and even an area of SiPM up to 1~cm$^2$ in one channel, corresponding to an input capacitance of approximately 5~nF. Thus, we conclude that KLauS5 can fulfill the TAO requirements for precise charge measurement. 
\end{abstract}


\setlength{\parindent}{2em}

\section{Introduction}
The Taishan Antineutrino Observatory (TAO, also known as JUNO-TAO) \cite{Abusleme:2020bzt} is a satellite experiment of the Jiangmen Underground Neutrino Observatory (JUNO) \cite{An:2015jdp,Djurcic:2015vqa}. It is proposed to precisely measure the reactor antineutrino spectrum via inverse beta decay (IBD) reactions based on a ton-scale gadolinium-doped liquid scintillator (GdLS) detector with a record energy resolution of $<$ 2\% at 1 MeV. The primary goal of TAO is to provide a reference spectrum for JUNO to eliminate the possible model dependence in the determination of neutrino mass ordering. Although a 3\%/$\sqrt{E}$ energy resolution is sufficient for TAO to serve as a reference detector of JUNO, the target energy resolution of TAO is capable of providing a benchmark of the reactor antineutrino spectrum to test nuclear databases. With a state-of-the-art detector, the additional scientific goals of TAO are to improve the nuclear physics knowledge of the neutron-rich isotopes in reactors, provide increased reliability in the measured isotopic antineutrino yields, search for sterile neutrinos, and verify the technology for reactor monitoring, safeguard, etc.

To reach the desired energy resolution of TAO, approximately 10-m$^2$ high-performance silicon photomultiplier (SiPM) \cite{Renker:2006ay} arrays are proposed to collect scintillation light with a coverage of $\sim$95\%. To mitigate the effect of SiPM dark noise, the SiPM arrays will be operated at -50~$^\circ$C to reduce the dark noise rate by approximately three orders of magnitude compared to that at room temperature \cite{Gola:2019idb}. To minimize the effects of the readout system of the SiPM arrays on the energy resolution, the system must precisely measure the charge triggered by photons at the single-photon level. An application-specific integrated circuit (ASIC) is one of the solutions for the SiPM readout system in TAO, which is designed to be operated in cold conditions and located close to the SiPM arrays, to achieve a good signal-to-noise ratio (SNR). The suitable ASICs for TAO must combine the features of single-photon detection, a 1 ns level time resolution, a high signal-to-noise ratio and low power consumption. The KLauS (Kan$\ddot{a}$le f$\ddot{u}$r die Ladungsauslese von Silizium-Photomultipliern) ASIC \cite{Dorn:2012zz}, developed by Heidelberg University, is found to be the closest to meet the requirements of the TAO readout system. It has 36 input channels and was originally designed for an analog hadron calorimeter (AHCAL) in the CALICE collaboration \cite{Andreev:2004uy}. Detailed characterizations were performed for the KLauS ASIC at room temperature in \cite{Briggl:2017kyj,Yuan:2019lub}; however, its performance at low temperature is still unknown, particularly at the TAO operating temperature of -50 $^\circ$C. In this work, we conduct detailed studies on evaluating the KLauS ASIC from room temperature to -50 $^\circ$C. The results of this work can not only provide guidance for TAO and other potential experiments to choose readout options but also provide essential feedback for KLauS developers to optimize the chip for applications at cryogenic temperatures.

This paper is organized as follows: we first introduce the TAO detector and the KLauS ASIC. Then, the testing setup used in this work is presented, together with a list of the key parameters of the KLauS ASIC to be tested. Finally, we report and discuss the performance of the KLauS ASIC measured at low temperatures with different inputs connected to the chip, i.e., charge injection circuits and different numbers of SiPMs.

\section{TAO detector and KLauS ASIC}
\begin{figure}[htbp]
	\centering
	\includegraphics[width=0.7\textwidth]{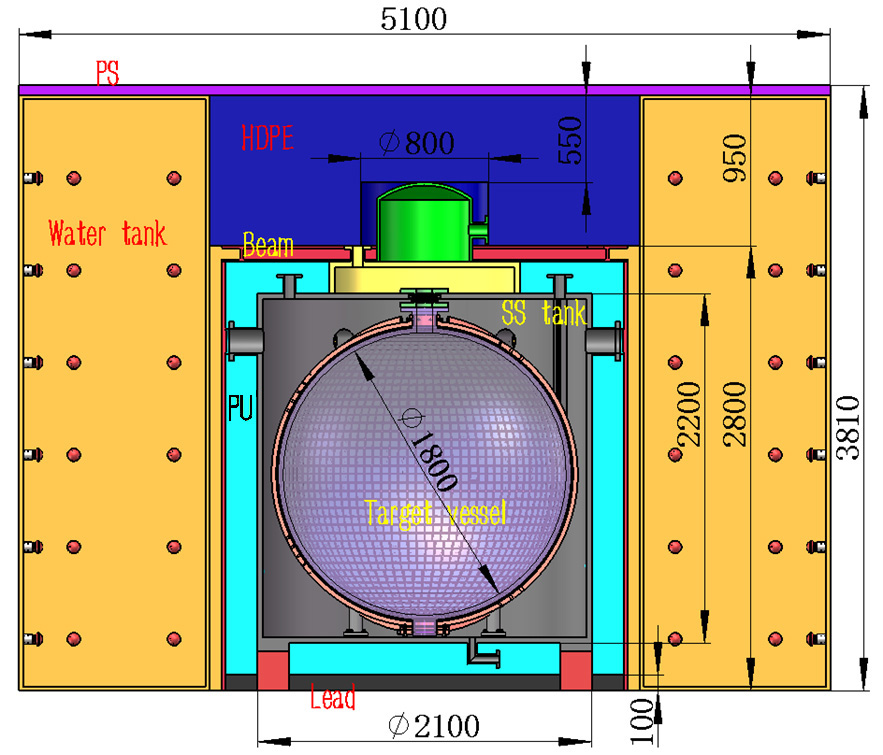}
\caption{Conceptual design of the TAO detector.The unit of the dimension is in milimeter.}
	\label{cdscheme}
\end{figure}

The TAO detector will be installed in a basement located outside of the reactor containment, approximately 30 meters from one of the reactor cores at the Taishan nuclear power plant. The conceptual design of the TAO detector was published in \cite{Abusleme:2020bzt}. It is composed of a central detector (CD) and two veto detectors, as shown in Figure~\ref{cdscheme}. The target material of the CD is GdLS with a total mass of 2.8 tons contained in an acrylic sphere with an inner diameter of 1.8 meters. The recipe of the GdLS is similar with that used in the Daya Bay experiment \cite{Ding:2008zzb}, but adapted for cold operation. Approximately 10-m$^2$ SiPMs will be installed on the inner surface of a copper shell to fully cover the acrylic sphere and collect scintillation light with sufficient light collection efficiency. The SiPMs are packaged in more than 4000 SiPM tiles. Each tile with dimensions of about 50~mm $\times$ 50~mm consists of 8$\times$8 SiPMs (6$\times$6~mm$^2$ for each SiPM). The coverage of the SiPM tiles is approximately 95\%, and the photon detection efficiency of the SiPMs must reach 50\%, which yields a photon detection of $\sim$4500 photoelectron (p.e.) per MeV. The copper shell and the SiPM tiles are immersed in a buffer liquid contained in a stainless steel (SS) tank. The CD will be cooled down via a refrigerator through cooling pipes deployed on the copper shell and the SS tank. It will be operated at -50 $^\circ$C to reduce the influences on the energy resolution from the SiPM dark noise. There are two readout options considered for the more than 4000 SiPM tiles in TAO. One is based on ASIC, and the other is based on discrete components. The discrete readout option is designed to connect all SiPMs in one tile for readout by a single channel. However, the ASIC-based readout option uses one chip with 36 inputs to readout one tile with 8X8 SiPMs, corresponding to two SiPMs in one channel for the KLauS ASIC, which allows us to have a high readout granularity at a level of 1~cm$^2$ per channel, corresponding to approximately 130k channels in total. A water Cerenkov detector surrounding the CD will be used to tag cosmic muons and shield the radioactive background from the basement. On top of the CD and the water Cerenkov detector, a plastic scintillator detector will be installed to tag muons.

\begin{figure}[htbp]
	\centering
	\includegraphics[width=0.7\textwidth]{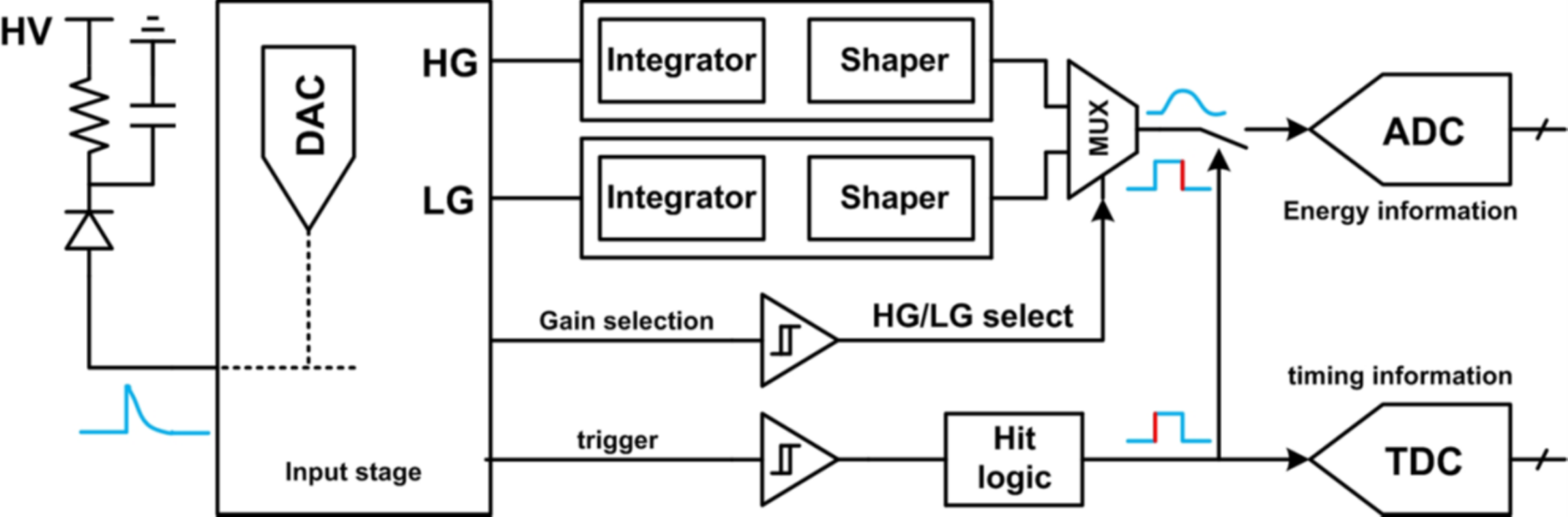}
\caption{Block level schematic diagram of an ASIC channel \cite{Briggl:2017kyj}.}
	\label{aa}
\end{figure}

The KLauS chip is a 36-channel ASIC fabricated in UMC 180-nm CMOS technology. A block level schematic diagram of one KLauS channel is shown in Figure~\ref{aa} \cite{Briggl:2017kyj}. The input stage buffers the detector current signal and distributes it to the two charge measurement branches and the two comparator branches. The SiPM bias voltage of each channel can be finely tuned in a 2~V range via an 8-bit digital-to-analog converter (DAC) implemented in the input stage. The current signal is integrated by a passive integrator and then shaped by an active shaping circuit. Charge integration is performed with the two charge measurement branches, known as the high gain scale (HG) and low gain scale (LG), corresponding to different charge conversion factors. The HG branch gives better equivalent noise charge (ENC) performance compared with the LG branch to ensure the measurement of single photoelectron (s.p.e.) spectra. Two charge conversion factors are available channel-wise in the HG branch, the so-called HG and medium gain scale (MG), which can be configured in the slow control. Similar to the HG branch, two scale factors are also configurable in the LG branch, the so-called LG and ultralow gain scale (ULG). The gain branch can be automatically selected by a gain-selection comparator with a configurable threshold. It determines which of the analog signals from the two gains is sampled and digitized by the following analog-to-digital converter (ADC). The detector current signal will also be compared with a predefined threshold in a time comparator controlled by a global 6-bit DAC for all channels and a channel-wise 4-bit DAC for fine tuning. The time comparator is used to record the time stamp of the hits and initiates the analog-to-digital conversion. The time stamp is provided by a high frequency clock counter combining the time-to-digital converter from the event trigger. The triggered analog signal is routed to an ADC that digitizes the analog information after a certain hold time (called the "hold-delay time") after the trigger time. The hold-delay time is configurable for each channel, and the hold-delay time can be optimized to yield the best SNR. A 10-bit successive-approximation-register (SAR) ADC is implemented in the KLauS ASIC for normal data taking, and a 12-bit pipeline ADC is also integrated on chip and can be used for the more precise readout of SiPMs. A digital control circuit initiates and controls the analog-to-digital conversion and passes the digitized data to the following digital part, which is responsible for combining, buffering, and sending the data to the data acquisition (DAQ) system. In this work, we use the fifth version of the KLauS chip (KLauS5) for testing, which has a time step of 25 ns. The latest version, the sixth version (KLauS6), has just been fabricated with an improved time resolution of 200 ps, with the other parts in KLauS6 remaining almost the same as those in KLauS5.

\section{Testing setup and key parameters to be tested}
\begin{figure}[htbp]
	\centering
	\includegraphics[width=0.8\textwidth]{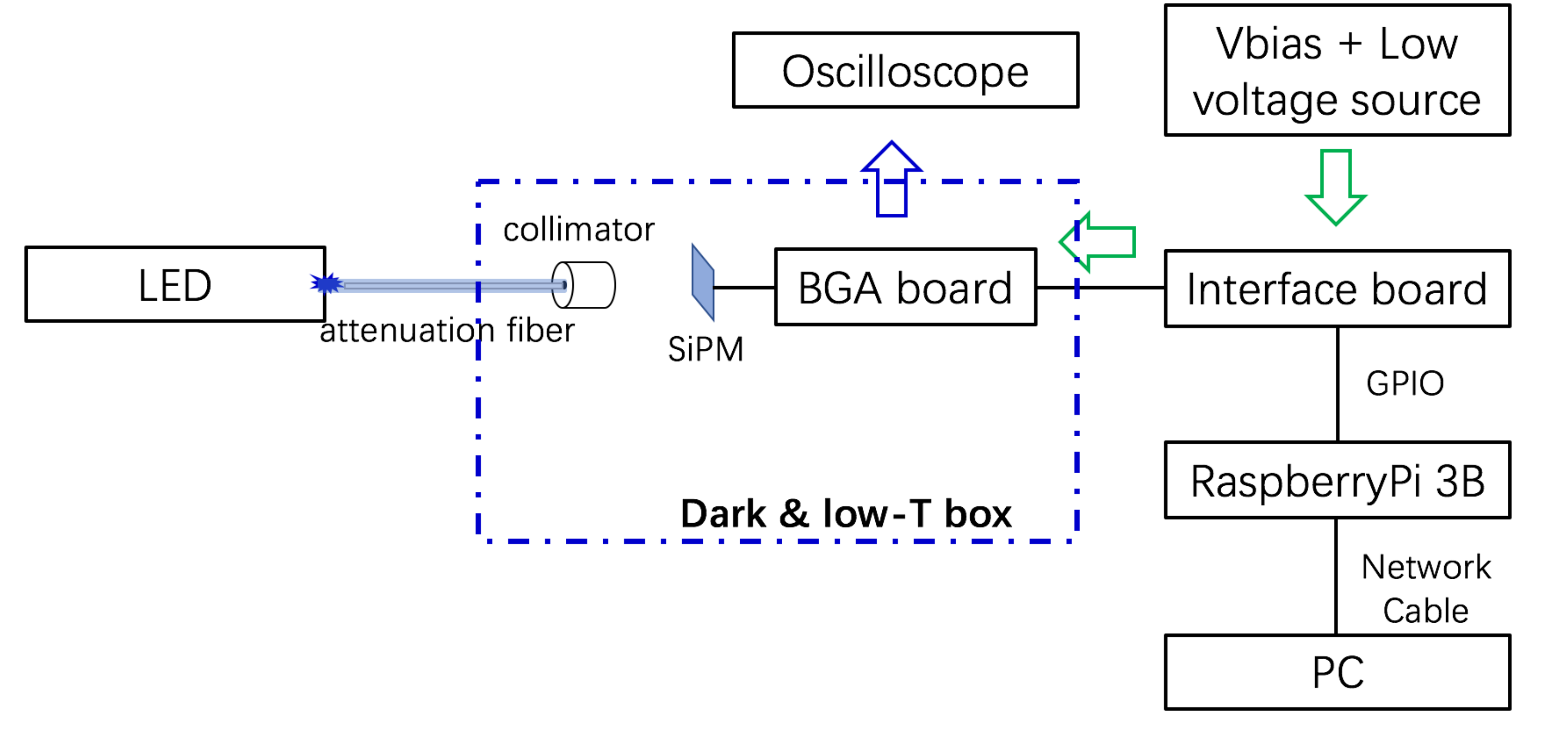}
\caption{Schematic drawing of the KLauS testing setup.}
	\label{a}
\end{figure}

A schematic drawing of the KLauS testing setup is shown in Figure~\ref{a}. The KLauS5 is bonded on an ASIC board located inside a high-low temperature test box, indicated by the blue dashed lines. The temperature inside the test box is adjustable within the range from -120 to 150~$^\circ$C. The test box also serves as a dark box to shield the SiPM samples from background illumination from outside of the test box. An LED, driven by a pulse generator, provides pulsed illumination to the SiPMs through an optical fiber integrated with a collimator. The SiPMs are connected to the input stage of the chip via an input connector on the ASIC board. An interface board, located outside of the test box, is used to provide power and a clock to the KLauS5 and provide a bias voltage to the SiPMs. The ASIC board and the interface board are connected via a ribbon cable. A raspberry Pi is connected to the interface board, which is used to configure the chip and take data. DAQ software, provided by the KLauS developers, is installed both in the raspberry Pi and a PC. The PC is connected to the raspberry Pi through a network cable so that the chip configuration and the data taking can be done with the PC. The analog signal before and after the shaping can be monitored by an oscilloscope. The SiPMs can be replaced with a charge injection circuit, which generates a known amount of charge to the KLauS5, so that part of the key parameters of the chip can be measured in a more efficient and easier way.

A few key parameters of the KLauS ASIC can significantly influence the performance of the TAO detector; therefore, they need to be carefully studied. These parameters are summarized as follows:
\begin{enumerate}
\item Charge noise. Ref.\cite{Yuan:2019lub} demonstrated that the ENC(Equivalent Noise Charge) is approximately 5~fC in the HG branch with an input capacitance of 33~pF. For the ASIC readout option in TAO, two SiPMs will be connected to one channel in parallel, corresponding to an input capacitance at the level of a few nanofarads. In TAO, the ENC must be less than 0.1 p.e. At this level, the contribution of the charge noise to the energy resolution becomes negligible compared to other factors.
\item Charge linearity. The linearity in charge measurements is one of the key parameters to evaluate and understand the energy nonlinearity of the TAO detector.
\item Gain uniformity among channels. Even though the gain of each readout channel can be calibrated, good uniformity among different channels significantly reduces the efforts in channel-wise configuration and simplifies the commissioning of the TAO detector.
\item Recovery time. TAO detects reactor antineutrinos via IBD, which is composed of a prompt signal from energy deposition of the positron and a delayed signal from the neutron capture on gadolinium. The time interval between the prompt and delayed signals is determined by the Gd neutron capture time of approximately 28~$\mu$s \cite{DayaBay:2012aa}. A recovery time of less than 1~$\mu$s is essential for TAO readout system to maintain high efficiency for IBD detection.
\item SNR. Because the SiPMs in the TAO detector mainly detect photons at the single-photon level, it is crucial for the readout system to have a high SNR to separate single-photon signals from the pedestals.
\item Power consumption. The power consumption of the KLauS5 is determined to be 3.3~mW per channel at room temperature, as reported in \cite{Yuan:2019lub}. In the TAO detector, since the SiPM readout system will be operated at -50 $^\circ$C, the baseline requirement of its total power dissipation is less than 1 kW to guarantee a stable and uniform temperature environment for operating the GdLS and SiPMs at -50~$^\circ$C.
\end{enumerate}

\section{KLauS5 characterization with charge injection} \label{section:3}
\subsection{Charge noise}
\begin{figure}[htb]
\begin{minipage}[ht]{0.5\linewidth}
   \centering
{\includegraphics[width=\linewidth]{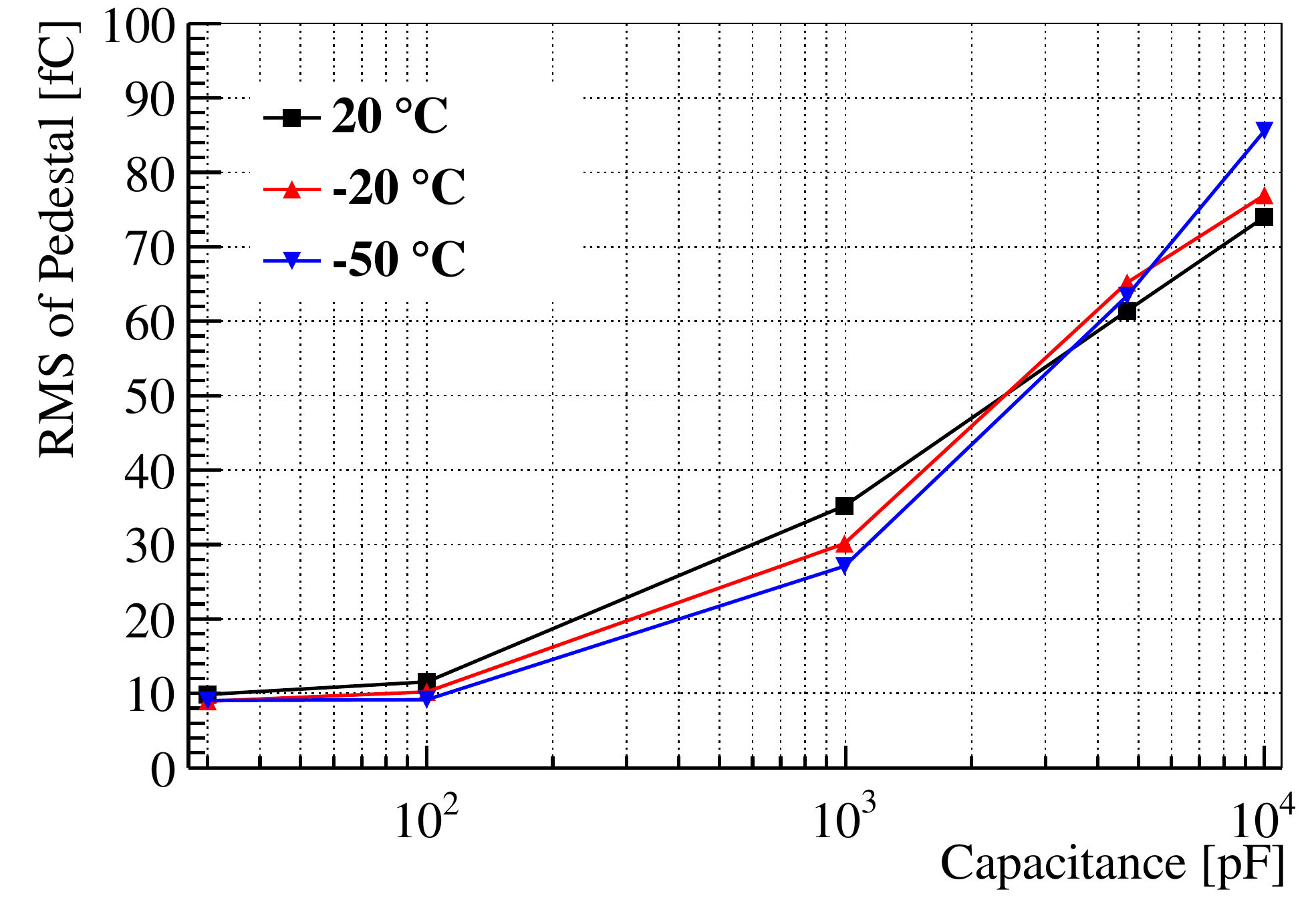}}{ (a)}
\end{minipage}
\hfill
\begin{minipage}[ht]{0.5\linewidth}
   \centering
{\includegraphics[width=\linewidth]{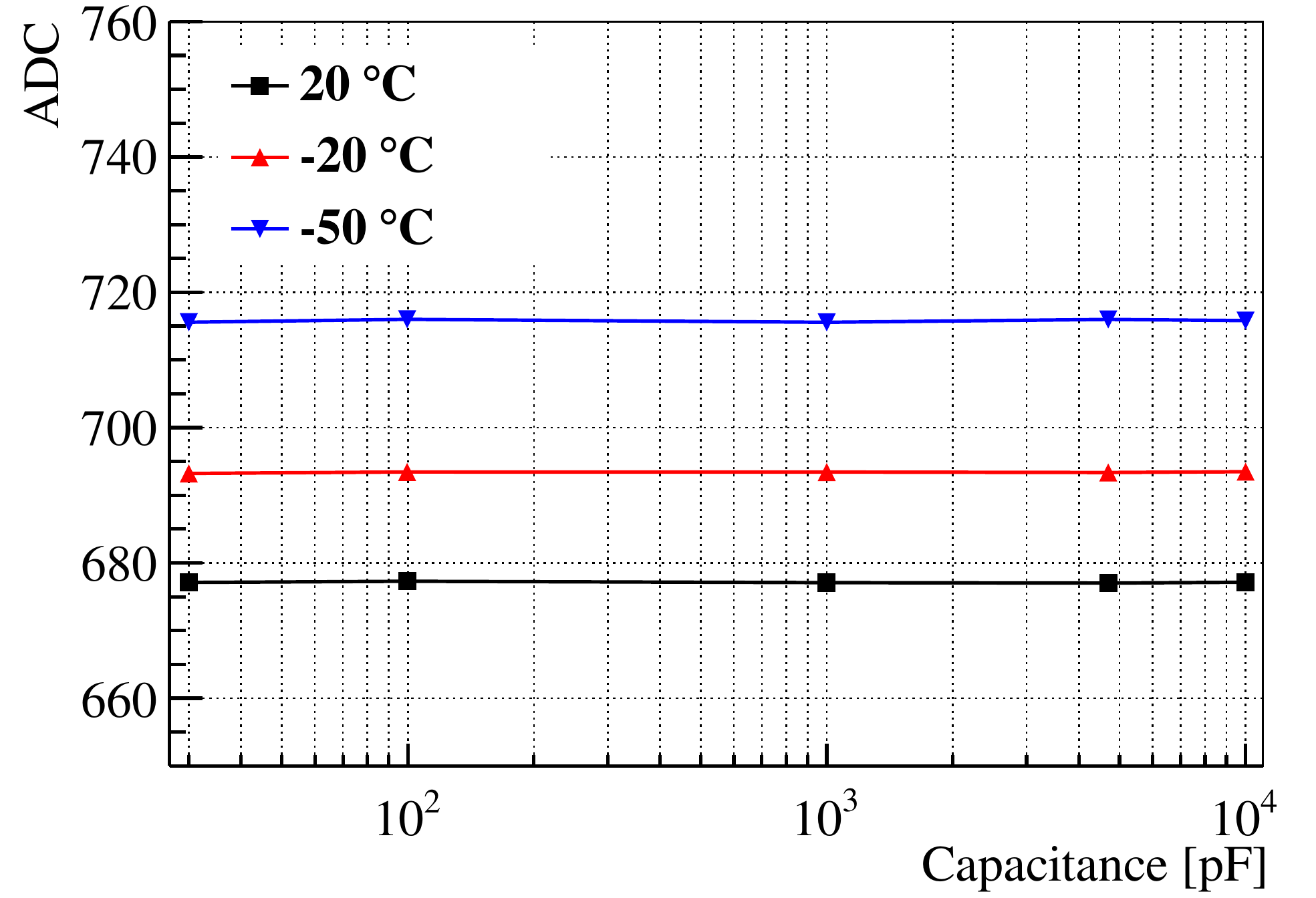}}{ (b)}
\end{minipage}
\caption{Charge noise (a) and positions of pedestals (b) as functions of input capacitance measured at three different temperatures of 20~$^\circ$C (black), -20~$^\circ$C (red) and -50~$^\circ$C (blue).}
   \label{bb}
\end{figure}

The noise performance is characterized by measuring the root mean square (RMS) value of the pedestal voltage. We measure the charge spectra of the baselines with the HG branch and the 10-bit SAR ADC by setting the trigger threshold to 0 in the time comparator. Therefore, the analog information sent to the ADC is purely triggered by the electronic noise. This measurement is repeated with different input capacitances and at different temperatures for one of the readout channels in the KLauS5. Figure~\ref{bb} (a) shows the RMS of the charge spectrum of the pedestal as a function of the input capacitance, measured at three different temperatures of 20~$^\circ$C (black), -20~$^\circ$C (red) and -50~$^\circ$C (blue) and was converted to fC with the gains of 13.02 ADC/fC@20~$^\circ$C, 11.97 ADC/fC@-20~$^\circ$C and 10.88 ADC/fC@-50~$^\circ$C. The ENC increases when the input capacitance increases. However, the temperature has a negligible impact on the ENC. The positions of the baselines, calculated from the mean values in their charge spectra, have a strong correlation with the temperature but do not depend on the input capacitance, as shown in Figure~\ref{bb} (b). At lower temperatures, the offset of the baseline becomes larger, which results in a smaller dynamic range. At -50~$^\circ$C, we observed 13\% degradation of the dynamic range compared with that at room temperature. However, this is not an issue for TAO since only a few photons are expected in each channel for the events of interest.

\subsection{Charge linearity and gain uniformity}
\begin{figure}[htb]
\begin{minipage}[ht]{0.5\linewidth}
	  \centering
{\includegraphics[width=\linewidth]{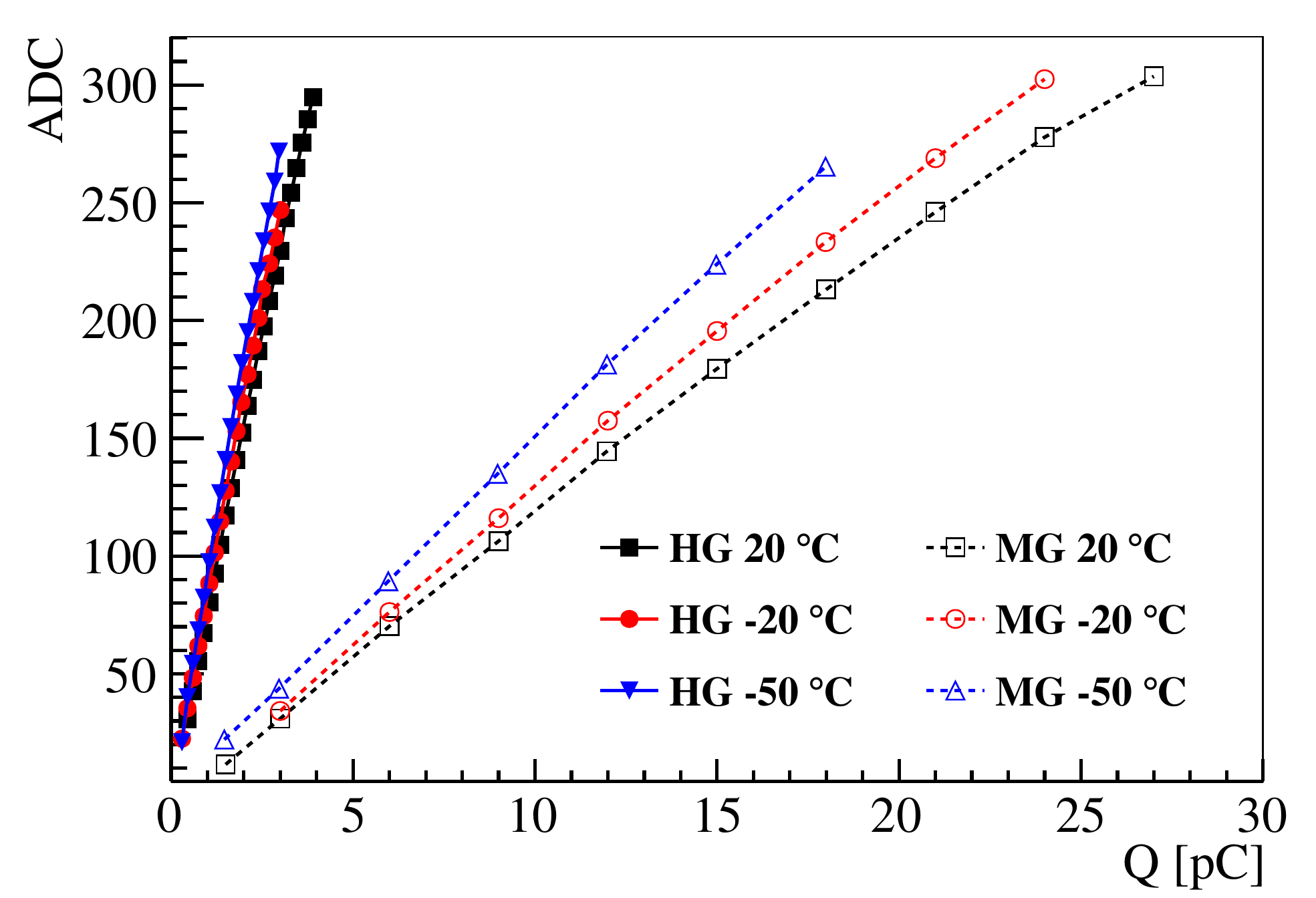}}{ (a)}
\end{minipage}
\hfill
\begin{minipage}[ht]{0.5\linewidth}
	  \centering
{\includegraphics[width=\linewidth]{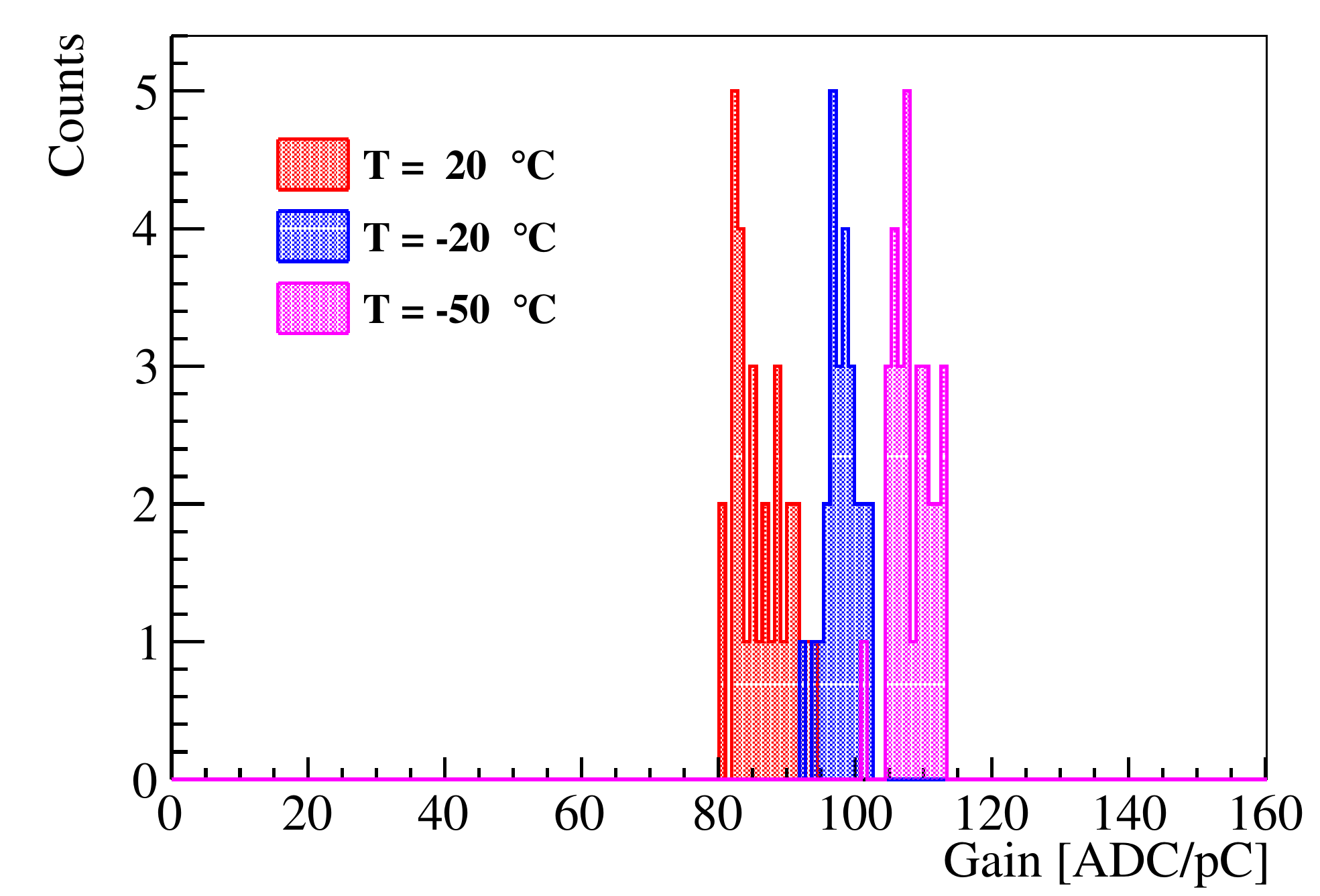}}{ (b)}
\end{minipage}
\caption{(a) Output ADC counts in the HG branch (solid lines) and MG branch (dashed lines) as a function of the input charge, measured at 20~$^\circ$C (black), -20~$^\circ$C (red) and -50~$^\circ$C (blue). (b) Gains of all 36 channels measured with the HG branch at temperatures of 20~$^\circ$C (red), -20~$^\circ$C (blue) and -50~$^\circ$C (pink).}
	  \label{c}
\end{figure}

Only the charge linearity in the HG and MG branches (of interest for TAO) are measured in this work. By injecting different amounts of charges into one of the channels in the KLauS5 ASIC, the ADC counts of the two branches are obtained, as shown in Figure~\ref{c} (a), measured with an input capacitance of 30~pF at 20~$^\circ$C (black), -20~$^\circ$C (red) and -50~$^\circ$C (blue). The pedestals have been subtracted in the ADC counts. The solid lines indicate the results measured with the HG branch, and the dashed lines represent those measured with the MG branch. When operating the KLauS5 from room temperature to -50~$^\circ$C, good linearity can be maintained in the most of the regions of the HG and MG branches. As shown in Table \ref{gain_linear_fit}, the fitting and quantified results of gain and goodness-of-fit were obtained with a simple linear fit. The dynamic ranges of both the HG and MG branches are significantly reduced at lower temperatures due to the aforementioned reason of a larger offset of the baseline. At -50~$^\circ$C, the HG branch can detect charges up to $\sim$4 pC; however, higher charges of $\sim$25~pC can be detected with the MG branch. If we assume that the gain of a SiPM is 2 $\times$ 10$^6$, corresponding to 0.32~pC, then the maximum number of photoelectrons that can be detected in each channel is approximately 12 for the HG branch and $\sim$75 for the MG branch, which can fulfill the dynamic range requirements (0-10 p.e.) for TAO. However, since the gain of SiPMs strongly depends on the SiPM type and applied bias voltage, optimization of the KLauS ASIC to better match it with the selected SiPMs would be achievable.

\begin{table}
    \centering
    \caption{Variation of gain and linearity with the fitting results at different temperatures}
    \begin{tabular}{|c|c|c|c|c|}\hline
    Temperature [$^{\circ}$C] & HG[ADC/pC] & Goodness-of-fit & MHG[ADC/pC] & Goodness-of-fit \\ \hline
    20  & 76.83  & 0.9992 & 11.6 & 0.9976 \\ \hline
    -20 & 83.56 & 0.9992 & 12.81 & 0.9988 \\ \hline
    -50 & 91.93 & 0.9988 & 14.85 & 0.9997 \\ \hline
    \end{tabular}
    \label{gain_linear_fit}
\end{table}

Figure~\ref{c} (b) shows the gains of all 36 channels measured with the HG branch. The gain of each channel is calculated from the slopes of the curves shown in Figure~\ref{c} (a). We can conclude that the gains of all channels increase by a factor of 30$\%$ from room temperature to -50~$^\circ$C. The gain uniformity among different channels is better than 10\% and consistent at different temperatures.

\subsection{Hold-delay time}
The hold-delay time is the time interval between the trigger time of the hit and the time at which analog information starts to be sampled and digitized. Ideally, the peak position of the analog signal after shaping should be digitized, and then, the maximum charge can be obtained, which results in the best SNR. The hold-delay time can be configured in the DAQ software via the 4-bit global DAC (gDAC) for all channels and the 4-bit fine DAC (fDAC) for individual channels. In this work, the hold-delay time is defined as 16 $\times$ gDAC + fDAC.

\begin{figure}[htb]
	\centering
	\includegraphics[width=1.\textwidth]{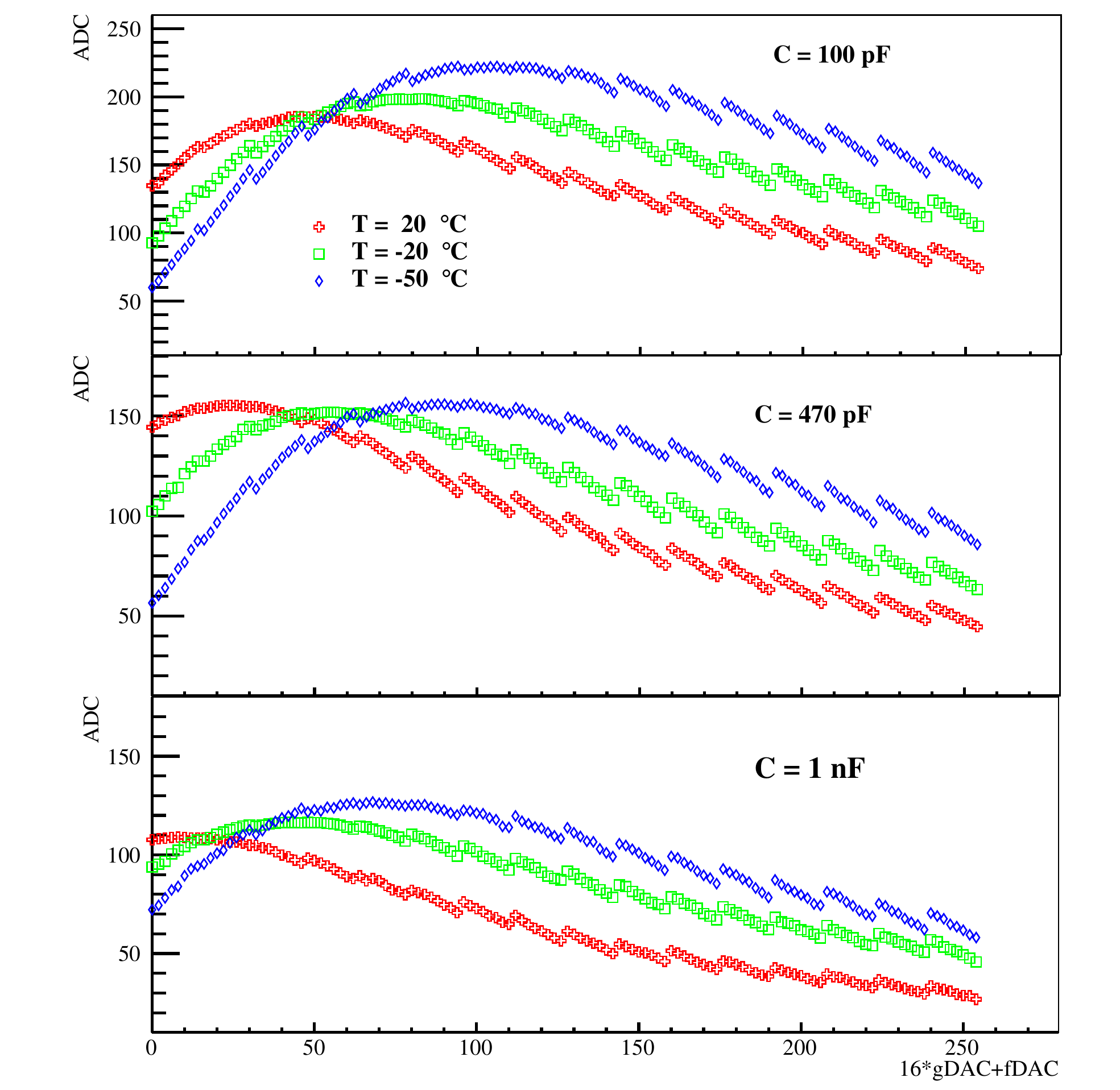}
\caption{Variation in the recorded ADC value for fixed charge injection with the hold delay, which is defined as 16 $\times$ gDAC + fDAC. The DAC represents that the time window is delayed to take the amplitude of the waveform from the trigger time.}
	\label{d}
\end{figure}

The hold-delay time is scanned within its full range for different input capacitances and input charges. The outputs of ADC counts are measured with the HG branch and the 10-bit SAR ADC. A fixed trigger threshold is used for all data points. The ADC counts as a function of the hold-delay time are shown in Figure~\ref{d}. The input capacitance is labeled on each plot, and the temperature is indicated with different colors. For a fixed input capacitance, the optimal hold-delay time increases when the KLauS5 operates at lower temperatures. When the input capacitance is increased, the optimal hold-delay time decreases for all tested temperatures and is even not achievable for the case of an input capacitance of 1~nF at room temperature. In TAO, the expected input capacitance is a few nF for each channel, which means that a small hold-delay time is preferred, which certainly might be out of the configurable range. Therefore, this could be optimized for readout of the large-area SiPMs and for operating the KLauS ASIC in cold conditions.

\subsection{Recovery time}
\begin{figure}[htb]
	\centering
	\includegraphics[width=0.8\textwidth]{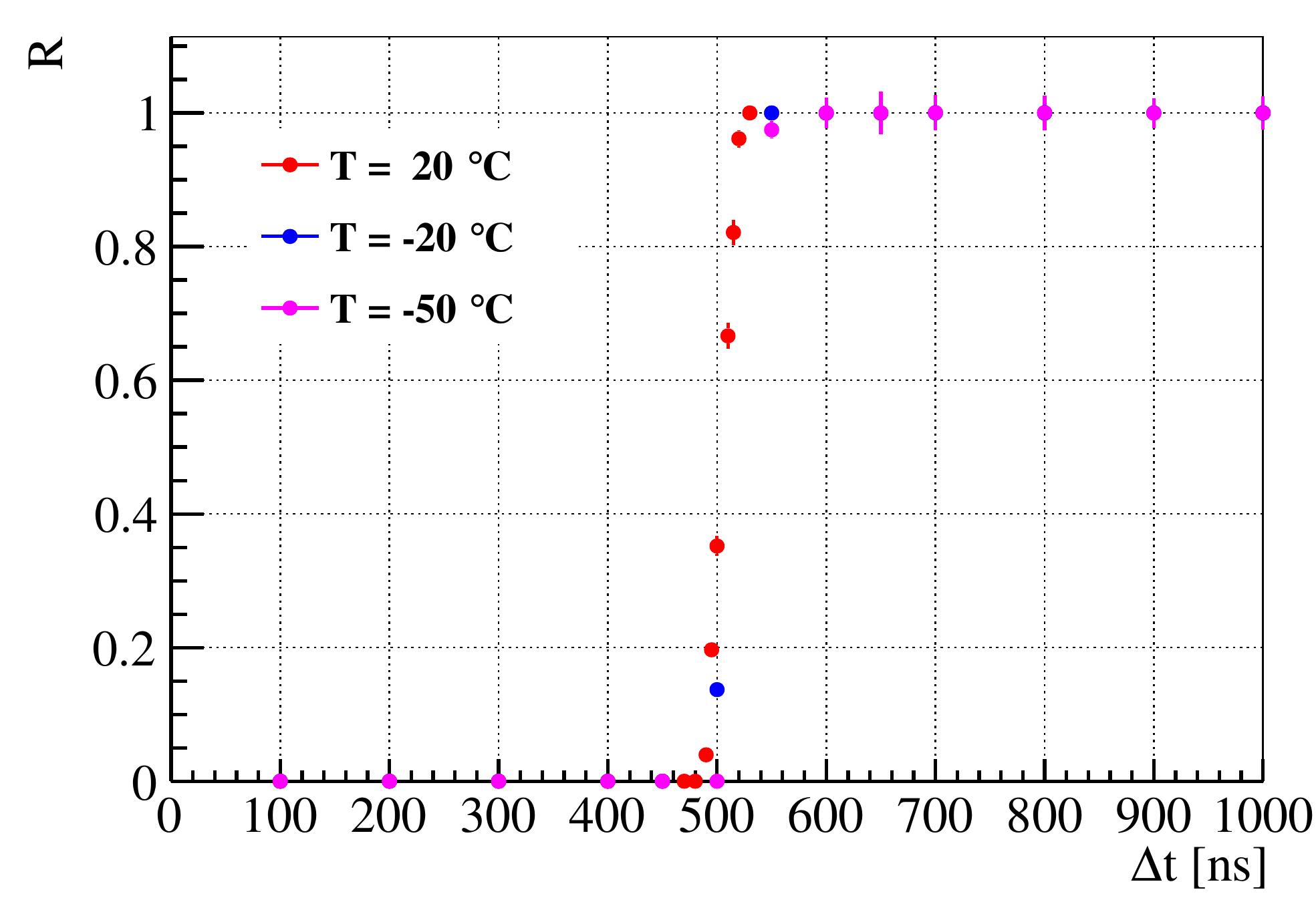}
\caption{Fraction of the second charge detected as a function of the time interval between the two injected charges at temperatures of 20~$^\circ$C (red), -20~$^\circ$C (blue) and -50~$^\circ$C (pink).}
	\label{e}
\end{figure}

The recovery time of the KLauS5, mainly stemming from the analog-to-digital conversion time and the sampling time, is studied at -50~$^\circ$C, -20~$^\circ$C and room temperature by directly injecting two pulses of charge with different magnitudes into the chip. The time interval between the two injected pulses is adjustable through the delay time of the second charge in the pulse generator. During the processing of the first pulse in the KLauS5 chip, the second pulse cannot be detected with 100\% efficiency. This feature is well demonstrated in Figure~\ref{e}, which shows the fraction of the second charge detected as a function of the time interval between the two injected pulses. We conclude that the KLauS5 chip fully recovers within 600 ns, which meets the requirements of IBD detection in TAO. The tested working temperatures have no significant impacts on the recovery time.

\subsection{Power consumption}
\begin{figure}[htb]
	\centering
	\includegraphics[width=0.5\textwidth]{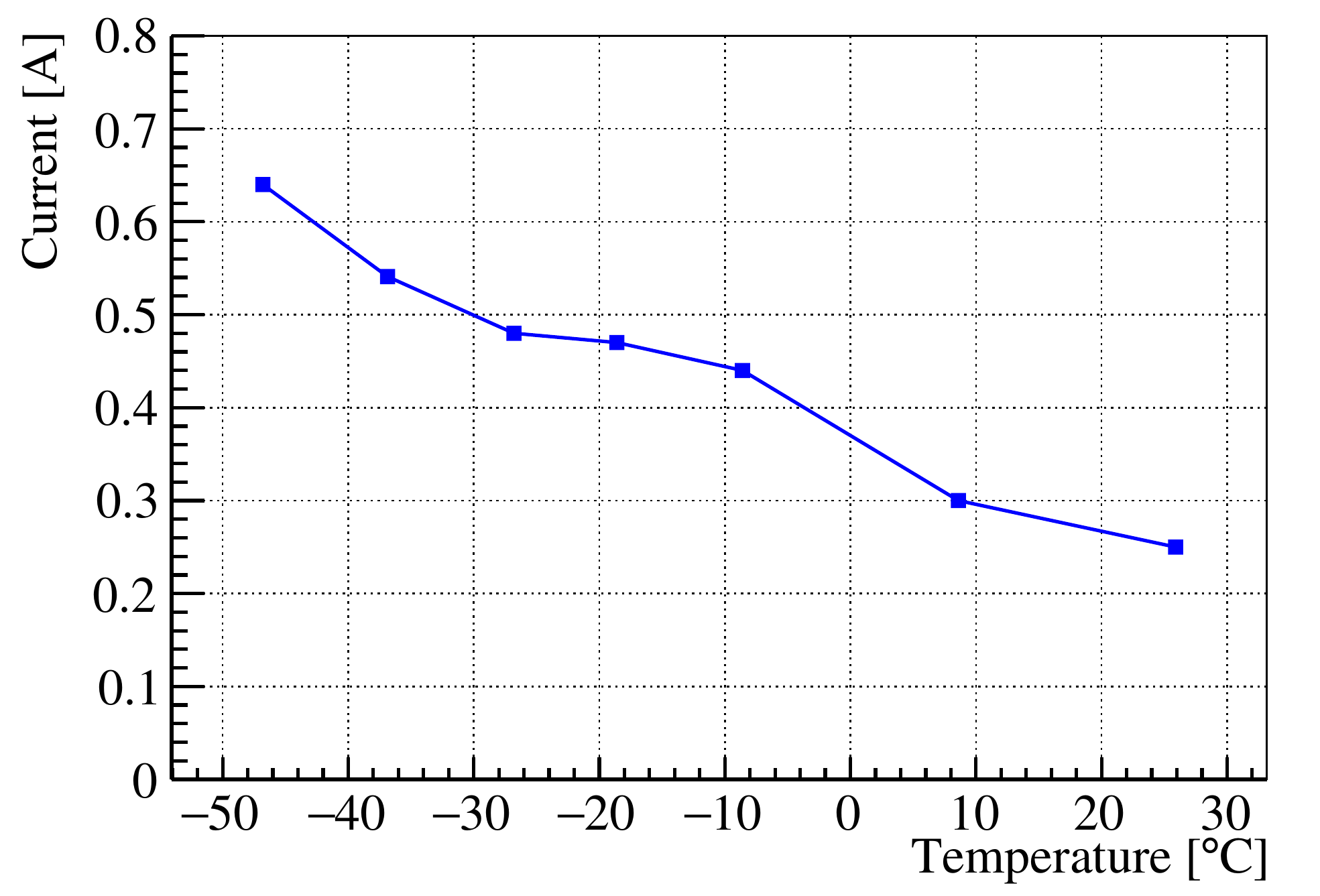}
\caption{Current read from the power supply used to provide power for the KLauS5 ASIC and the interface board as a function of the working temperature.}
	\label{b}
\end{figure}

The power consumption of the KLauS5 ASIC was measured to be approximately 3.3~mW per channel at room temperature in \cite{Yuan:2019lub}. In this work, the power consumption of the KLauS5 is not directly measured; however, the current of the power supply used for both the ASIC board and the interface board is monitored at different temperatures, as shown in Figure~\ref{b}. The current read from the power supply at -50 $^\circ$C  is 3 times higher than that at room temperature. This indicates that the total power consumption of the BGA board and the interface board increases by a factor of 3. If we assume that this factor only stems from the KLauS5, then conservatively, the power consumption of the KLauS5 ASIC will be higher by up to 3 times, which results in a total power consumption of less than 1.3~kW in the TAO detector. This number does not fulfill the baseline requirement of 1~kW, but it is still tolerable for TAO. Furthermore, the power consumption can be further reduced by optimizing the KLauS dedicated to operation at low temperatures.

\section{KLauS5 characterization with SiPMs}
\label{section:4}
A SiPM array manufactured by Hamamatsu is used for evaluation of the KLauS5. The model number of the SiPM array is S13371-6050CQ-02 \cite{hpk}. It consists of 2$\times$2 independent SiPMs. Each SiPM is 6$\times$6~mm$^2$ and packaged in a ceramic frame with two pins connected to the cathode and anode of the SiPM. The key parameters of the SiPM array relevant to this work are listed in Table~\ref{bs2} \cite{hpk}. One or more SiPMs are connected in parallel to one of the channels in the KLauS5, corresponding to the input capacitance changing from 1.2~nF to 4.8~nF. In the following discussions, we only report the measurements made at -20~$^\circ$C and -50~$^\circ$C since at room temperature, the baselines are heavily distorted because of the extremely high dark noise rates of the SiPMs and the signals of s.p.e. can only be observed with one SiPM connected to the chip.

\begin{table}
	\caption{Key parameters of the SiPM array at 25~$^\circ$C}
	\vspace{8pt}
	\centering
	\begin{tabular}{cc}
		\hline
		Number of Channels & 4 (2$\times$2) \\
		Active Area &  5.96 $\times$ 5.85 mm$^2$ \\
		Pixel Pitch & 50 $\mu$m \\
		Number of pixels & 13,923\\
		Break Voltage&	53 $\pm$ 3 V \\
		Gain &	2.55$\times$10$^6$ at an overvoltage of 4 V 	\\
		DCR & $\sim$4.0 Mcps/ch at an overvoltage of 4 V\\
		Terminal Capacitance/ch. & 1200 pF \\
		\hline       
	\end{tabular}
	\label{bs2}
\end{table}

\begin{figure}
	\centering
	\includegraphics[width=0.8\textwidth]{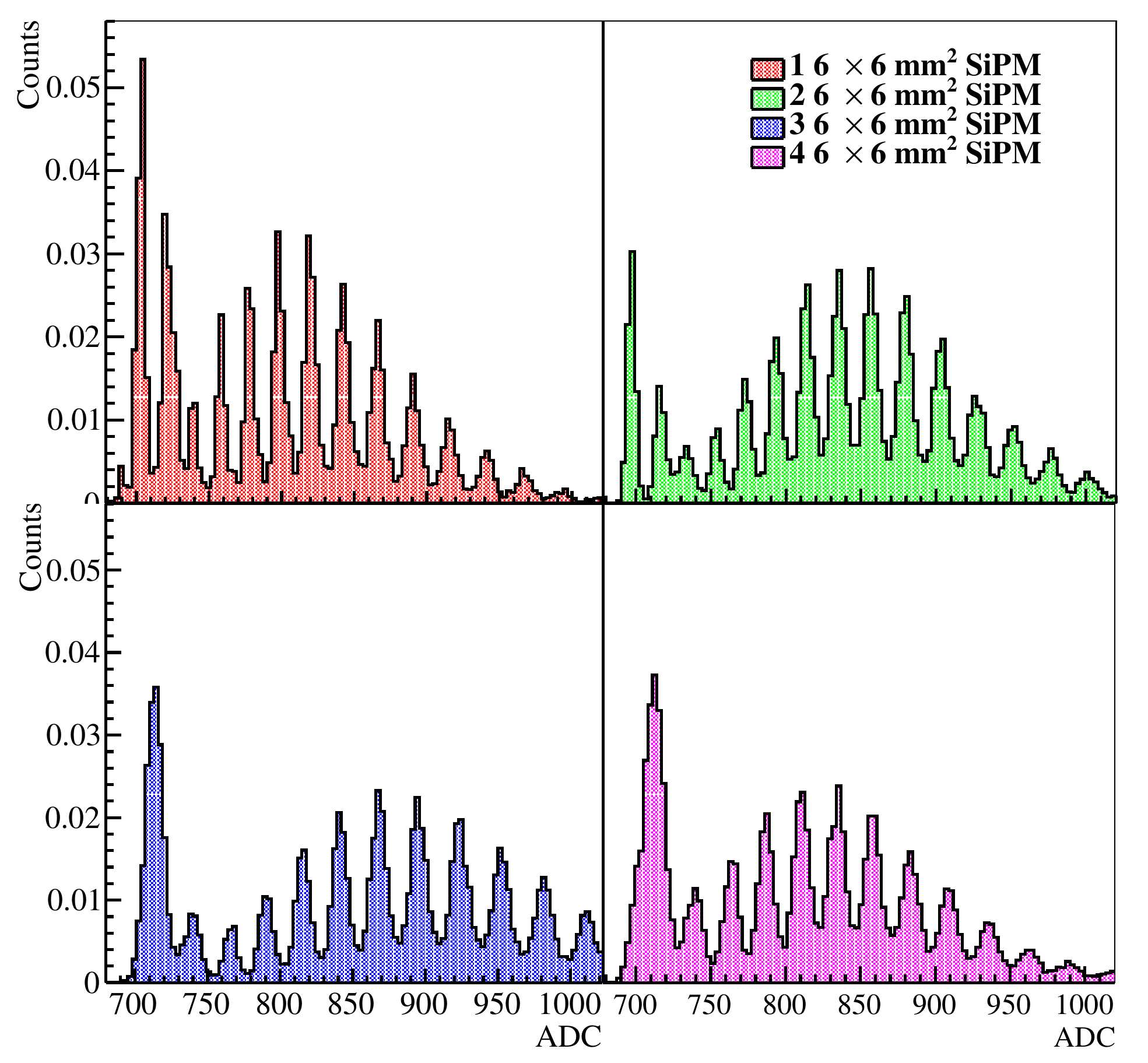}
\caption{Charge spectra measured by the KLauS5 in the HG branch at -50~$^\circ$C, with one (red), two (green), three (blue) and four (pink) SiPMs connected to the chip. The overvoltage is 1.4~V, 1.4~V, 1.7~V and 1.4~V, respectively.}
	\label{gg}
\end{figure}

The charge spectra are measured with the HG branch and the 10-bit SAR ADC with pulsed light illumination of SiPMs at the two temperatures. According to the measured charge spectra, the intensity of the light is optimized to well match the dynamic range of the HG branch by tuning the voltage applied to the LED light source. Figure~\ref{gg} shows typical charge spectra taken at -50~$^\circ$C with an overvoltage of about 1.4 V, in which the number of detected photons can be well distinguished. The KLauS5 chip shows excellent performance at -50~$^\circ$C for the tested SiPMs even at the small overvoltage of about 1.3~V and with area up to 1.4~cm$^2$, corresponding to a gain of about 1$\times$10$^6$ and the input capacitance of approximately 5~nF, respectively. The four plots in Figure~\ref{gg} correspond to one, two, three and four SiPMs connected in parallel to one of the channels in the KLauS5. The first peak in each plot is the signal of s.p.e. triggered by the SiPM dark noise, and the subsequent peak is caused by optical cross-talk, which is one of the typical features of SiPMs \cite{Acerbi:2019qgp}. The remaining peaks are mainly triggered by the incident light, and the number of detected photons follows a Poisson distribution. The charge spectra measured at -20~$^\circ$C show features similar to those in Figure~\ref{gg}.

\begin{figure}
\begin{minipage}[ht]{0.5\linewidth}
	  \centering
{\includegraphics[width=\linewidth]{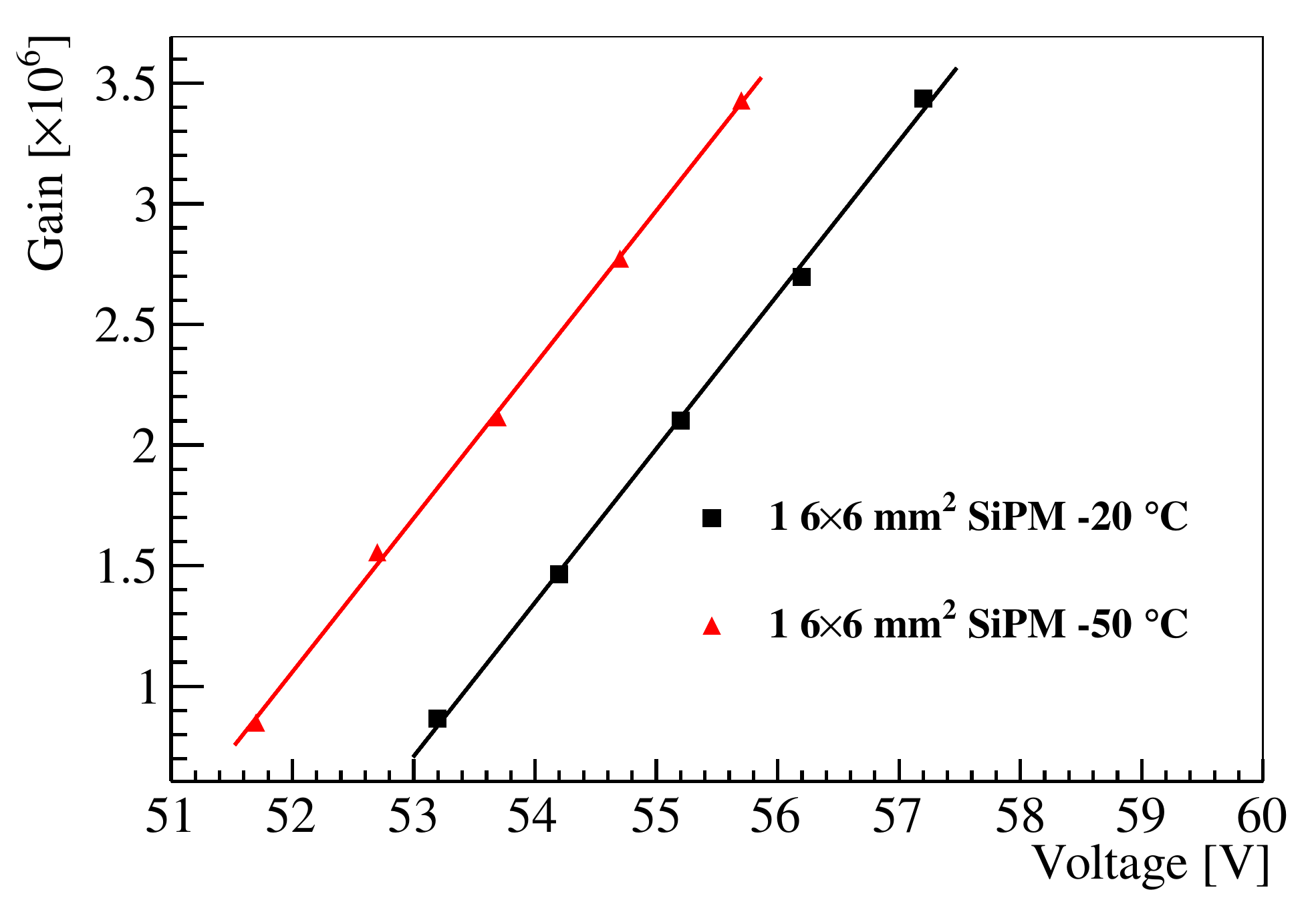}}{ (a)}
\end{minipage}
\hfill
\begin{minipage}[ht]{0.5\linewidth}
	  \centering
{\includegraphics[width=\linewidth]{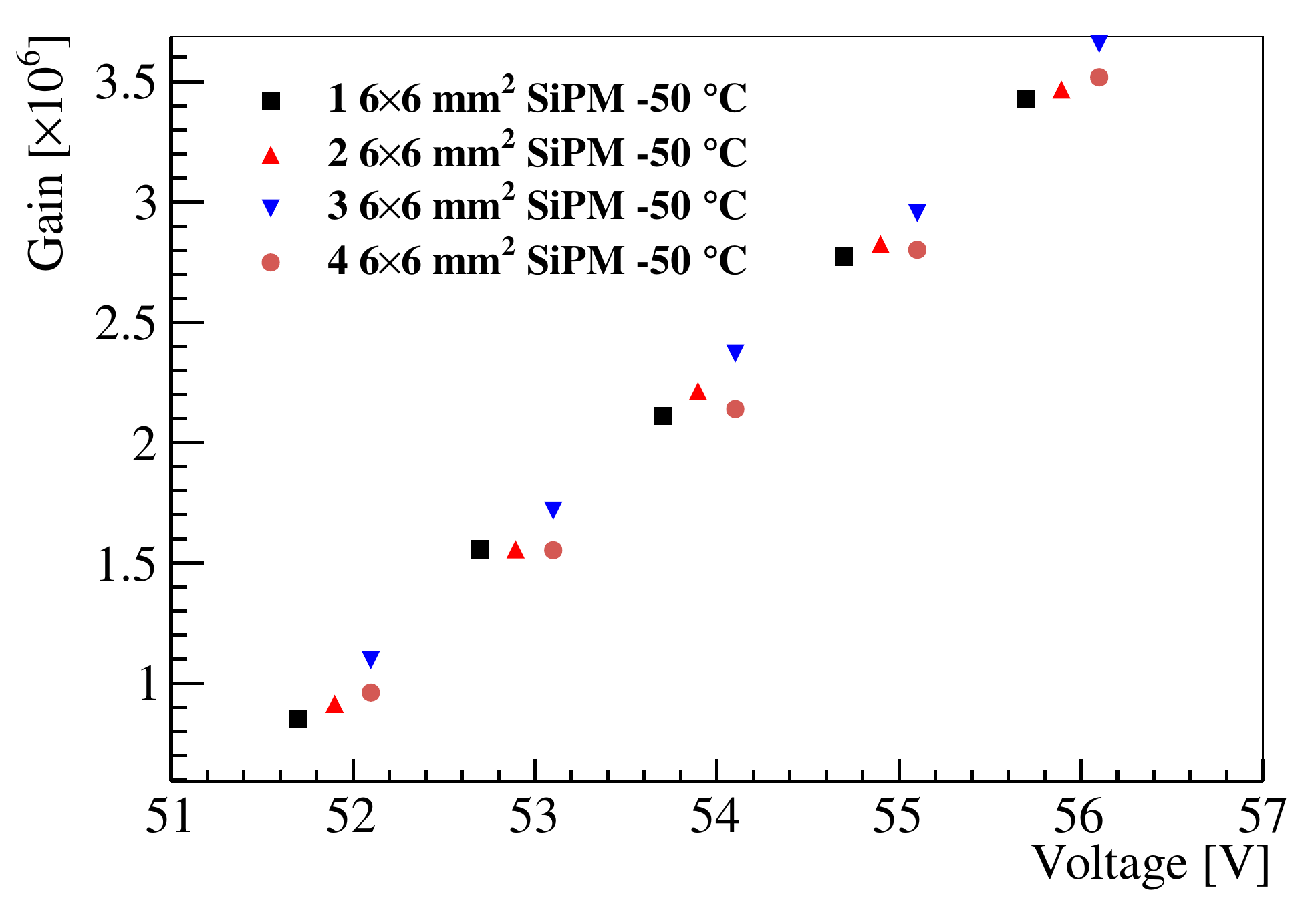}}{ (b)}
\end{minipage}
\caption{(a) Gain of the SiPM as a function of the bias voltage at -20~$^\circ$C (black) and -50~$^\circ$C (red) for a single SiPM, fitted with first-order polynomial functions. (b) Gain of the SiPMs as a function of the bias voltage at -50~$^\circ$C when connecting one (black), two (red), three (blue) and four (brown) SiPMs to one channel in the KLauS5.}
	  \label{led_PE_gain}
\end{figure}

The gain of the SiPMs can be extracted by fitting to the charge spectra shown in Figure~\ref{gg}, which is determined by the average distance between the two adjacent peaks. The gains as functions of the bias voltage are shown in Figure~\ref{led_PE_gain}. In Figure~\ref{led_PE_gain} (a), only one SiPM is connected to the KLauS5, and the gains are measured at -20~$^\circ$C (black line) and -50~$^\circ$C (red line). In Figure~\ref{led_PE_gain} (b), the gains are measured only at -50~$^\circ$C, with one or more SiPMs connected to the chip. The breakdown voltages are estimated by fitting the data points in Figure~\ref{led_PE_gain} with first-order polynomial functions demonstrated in Figure~\ref{led_PE_gain} (a), which are summarized in Table~\ref{work_V}. This shows that the breakdown voltages of connecting different numbers of SiPMs are consistent within 0.8\% at the tested temperatures.

\begin{table}
    \centering
    \caption{Breakdown voltages of one and more SiPMs connected in parallel at the two temperatures.}
    \begin{tabular}{|c|c|c|c|c|}\hline
    Temperature [$^{\circ}$C]$\backslash$ Breakdown voltage [V]  & 1 6x6mm$^2$ & 2 6x6mm$^2$ & 3 6x6mm$^2$ &4 6x6mm$^2$ \\ \hline
    -20  & 51.9  & 52.0  & 51.9 & 51.8 \\ \hline
    -50 & 50.3 & 50.5 & 50.4 & 50.7 \\ \hline
    \end{tabular}
    \label{work_V}
\end{table}

\begin{figure}
\begin{minipage}[ht]{0.5\linewidth}
	  \centering
{\includegraphics[width=\linewidth]{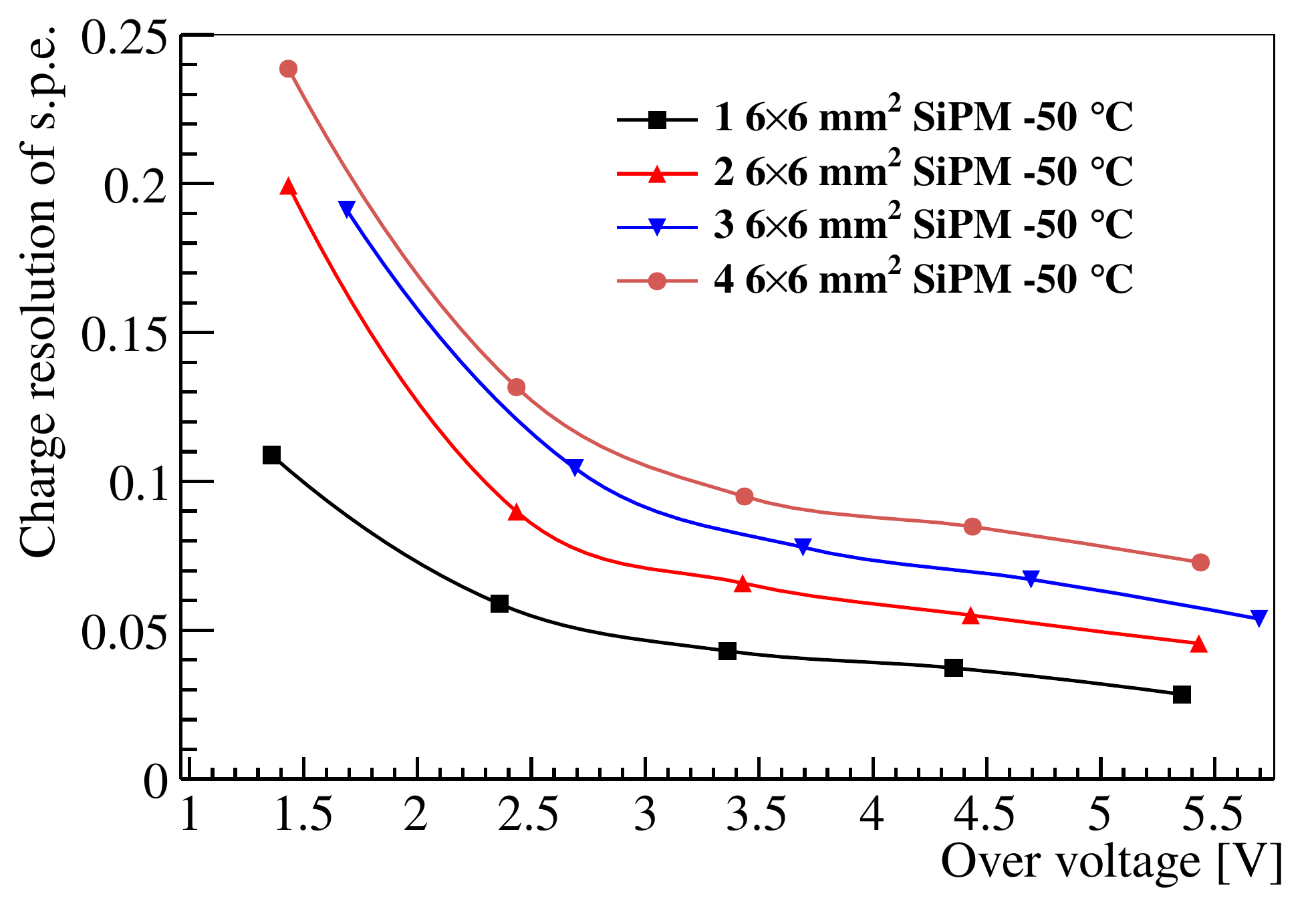}}{ (a)}
\end{minipage}
\hfill
\begin{minipage}[ht]{0.5\linewidth}
	  \centering
{\includegraphics[width=\linewidth]{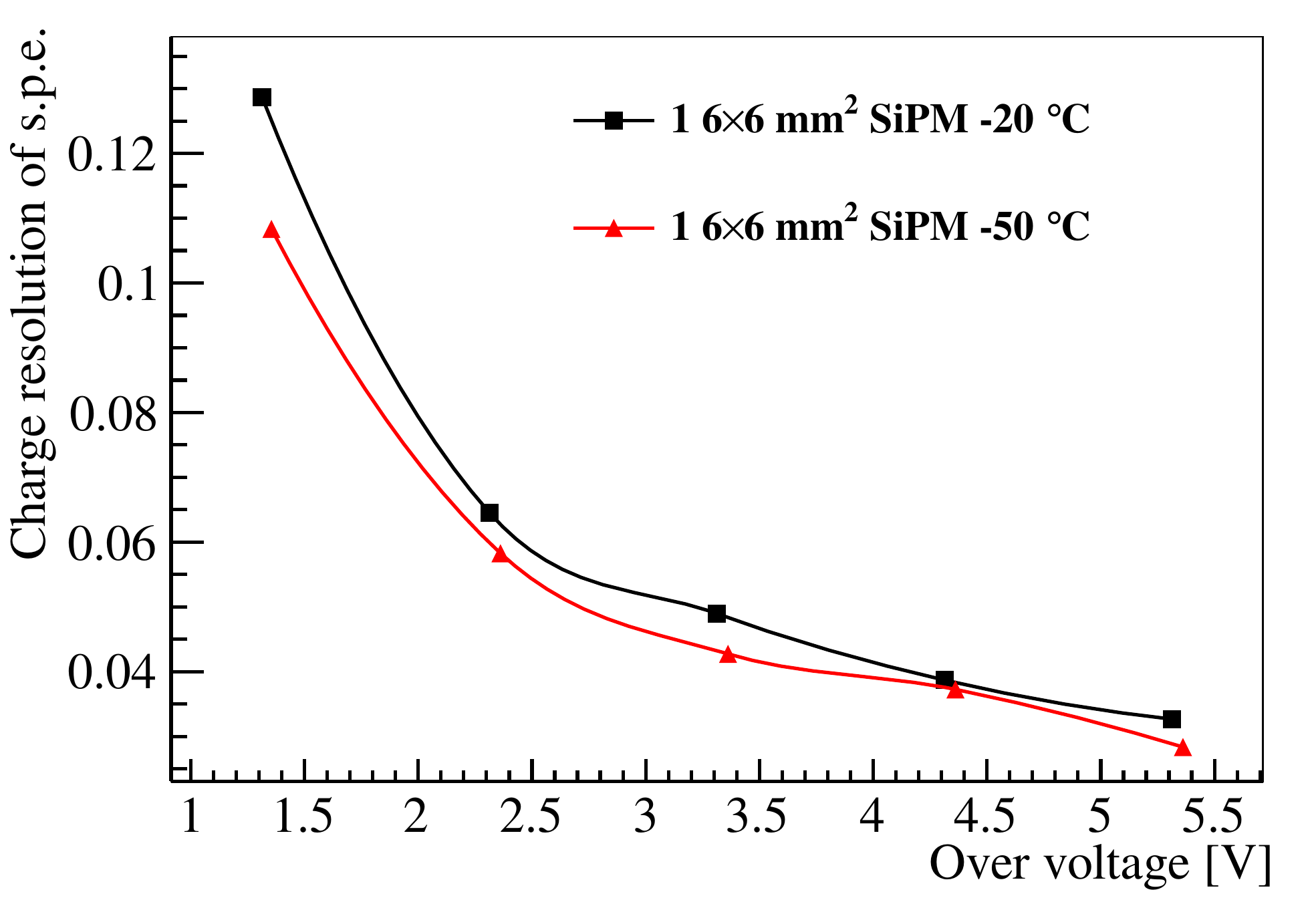}}{ (b)}
\end{minipage}
\caption{(a) Charge resolution of s.p.e. as a function of the overvoltage at -50~$^\circ$C with one (black), two (red), three (blue) and four (brown) SiPMs connected. (b) Charge resolution of an s.p.e. as a function of the overvoltage at -20~$^\circ$C (black) and -50~$^\circ$C (red) for a single SiPM.}
	  \label{single_photon_resolution}
\end{figure}

Because of the larger input capacitance, the charge resolution of s.p.e. deteriorates when more SiPM cells are connected. This feature is well demonstrated in Figure~\ref{single_photon_resolution} (a), which shows the charge resolution of s.p.e. as a function of the overvoltage when connecting different numbers of SiPMs. The charge resolution of s.p.e. is defined as the ratio of the standard deviation of the single-photon peak to the gain, which can be achieved by fitting to the measured charge spectrum. The spread of the breakdown voltage of the tested SiPMs is another reason for the worse resolution when more cells are connected in parallel. At -50~$^\circ$C, the charge resolution is better than 15\% when the overvoltage is larger than 2 V (an approximately 1.5$\times$10$^6$ gain), even with the input of 4 SiPMs. The lower temperature can help improve the resolution due to the lower dark noise rate, as shown in Figure~\ref{single_photon_resolution} (b), where the resolution is compared and presented at -50~$^\circ$C and -20~$^\circ$C at different overvoltages for a single SiPM.

\begin{figure}
\begin{minipage}[ht]{0.5\linewidth}
	  \centering
{\includegraphics[width=\linewidth]{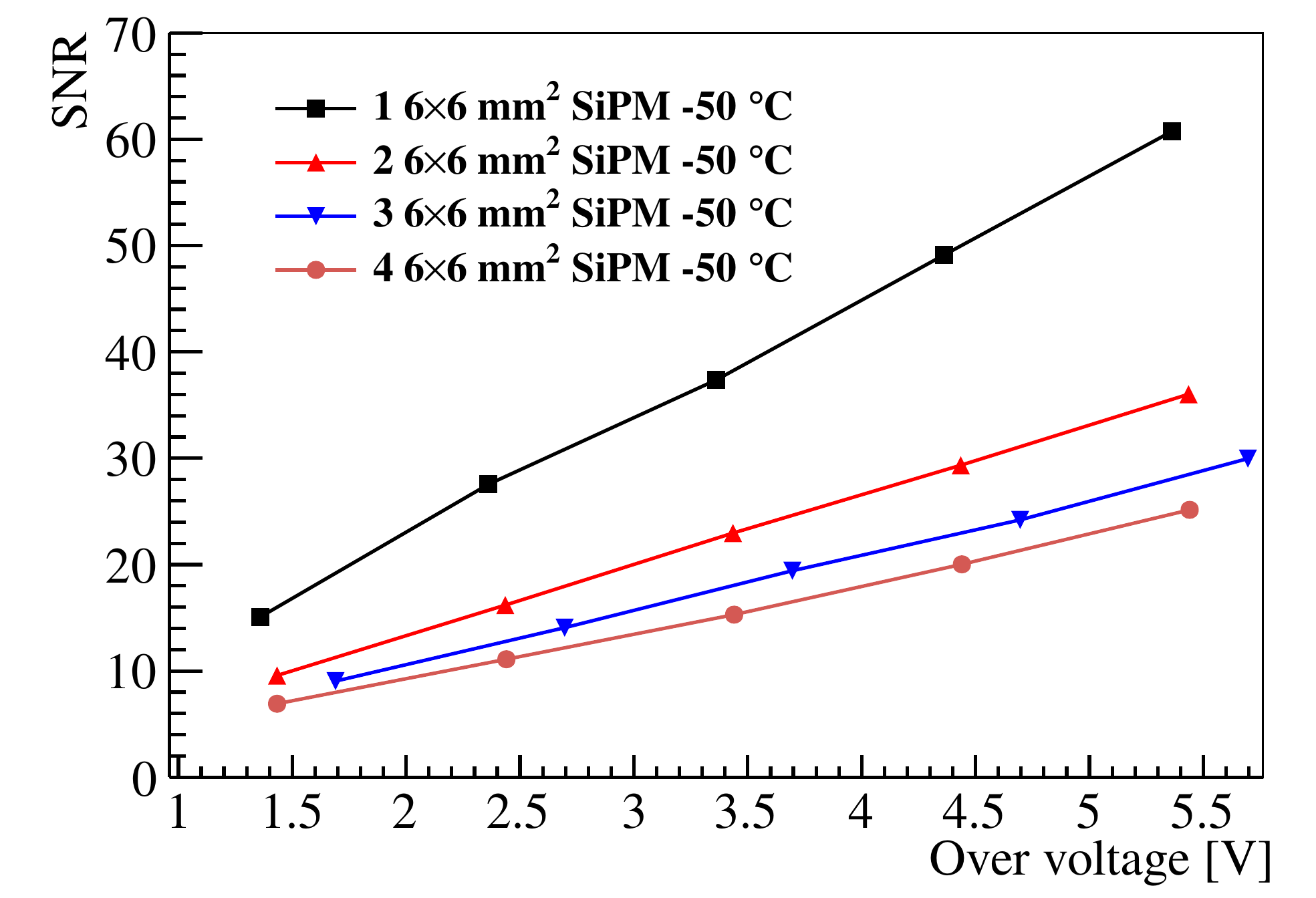}}{ (a)}
\end{minipage}
\hfill
\begin{minipage}[ht]{0.5\linewidth}
	  \centering
{\includegraphics[width=\linewidth]{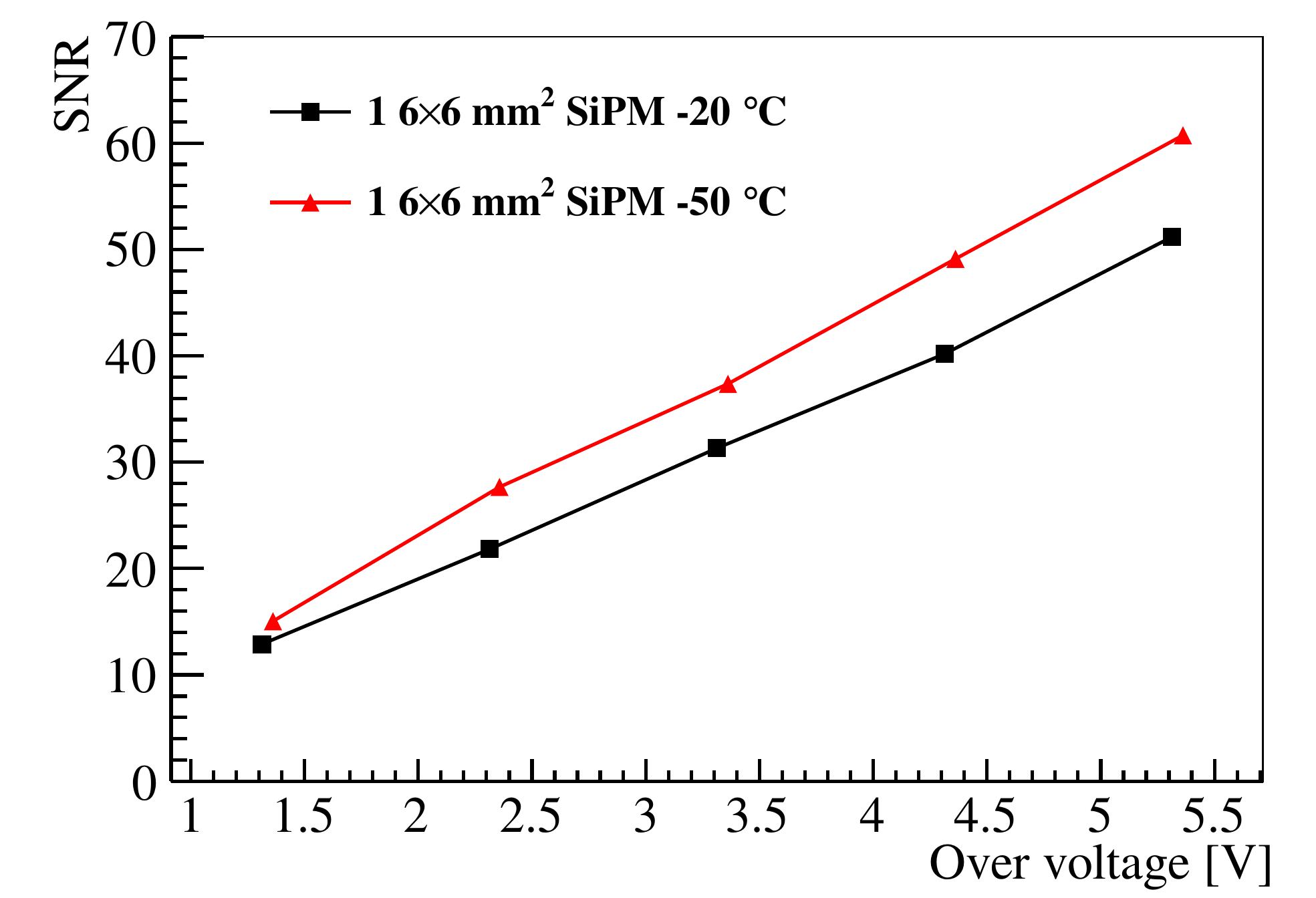}}{ (b)}
\end{minipage}
\caption{(a) SNR as a function of the overvoltage at -50~$^\circ$C with one (black), two (red), three (blue) and four (brown) SiPMs connected to the chip. (b) SNR as a function of the overvoltage at -20~$^\circ$C (black) and -50~$^\circ$C (red) with the input of a single SiPM.}
	  \label{SNR}
\end{figure}

The SNR is defined as the ratio between the gain and the standard deviation of the pedestal. The pedestal is measured by setting the trigger threshold to 0 in the time comparator and reducing the SiPM bias voltage to below its breakdown voltage. Thus, the chip can be purely triggered by noise. The SNRs are calculated for each measured data set and shown in Figure~\ref{SNR} as a function of the overvoltage. From Figure~\ref{SNR} (a), we can conclude that the SNR becomes worse when more SiPM cells are connected to one channel; however, it is better than 10 with overvoltages of larger than 2~V even for an input SiPM area up to 1~cm$^2$. Figure~\ref{SNR} (b) shows that the SNR can be improved by a factor of approximately 15\% at -50~$^\circ$C compared to that at -20~$^\circ$C, because of less dark noise rate.

In general, the KLauS5 ASIC shows an excellent performance at -50~$^\circ$C. The gain and capacitance of the SiPMs are the two major factors that can significantly impact the performance of the KLauS chip. In TAO, a gain of 1$\times$10$^6$ (the tested gain in this work) can be achieved easily since a large bias voltage is preferred to guarantee a high photon detection efficiency of 50\%. However, the bias voltage cannot be too high to keep the probability of correlated avalanches at an acceptable level. The terminal capacitance of the SiPMs must be minimized to enhance the SNR of KLauS. On the market, several SiPMs from different vendors can fulfill the aforementioned requirements, so the KLauS chip is a good candidate for the readout of large-area SiPMs.

\section{Conclusion}
TAO is proposed to precisely measure the reactor antineutrino spectrum with a record energy resolution of less than 2\% at 1 MeV, based on a 2.8 ton GdLS detector. Approximately 10 m$^2$ high-performance SiPMs, operated at -50~$^\circ$C, are proposed to collect scintillation light with sufficient light collection efficiency. The readout system of the SiPMs must measure the charge with good precision at the single-photon level to guarantee that its influence on the energy resolution is negligible. Meanwhile, it should also meet the requirements for timing and operating in cold conditions. ASICs are interesting readout solutions for TAO, among which the KLauS ASIC, developed by Heidelberg University, shows excellent performance at room temperature and is of interest for TAO. In this work, we carefully characterized the KLauS5 ASIC from room temperature to low temperatures, particularly at -50~$^\circ$C. The results show that KLauS5 can work normally down to -50~$^\circ$C, and no significant changes are observed for the charge noise, charge linearity, gain uniformity among channels and recovery time. Both the resolution of s.p.e and the SNR can fulfill the TAO requirements with the gain of the SiPMs greater than 1.5$\times$10$^6$, even for the case of an input SiPM area up to 1~cm$^2$ in one channel. Based on a conservative estimation, the power consumption of the chip is higher by up to 3 times compared with that at room temperature, which goes beyond the TAO baseline requirement, but still affordable. Generally, the existing KLauS5 ASIC can meet the TAO requirements for charge measurement, and the next version of KLauS (KLauS6, available now) with a better timing resolution of 200 ps can meet the timing requirement of 1~ns and is expected to be a good candidate for the TAO readout system. Moreover, the KLauS ASIC can be further improved to make it more suitable for operation in cold conditions, such as by optimizing the power consumption, hold-delay time, charge noise, etc. 

\section*{Acknowledgment}
We gratefully acknowledge support from National Natural Science Foundation of China (NSFC) under grant No. 11875279 and from CAS-IHEP Fund for PRC$\backslash$US Collaboration in HEP. And we also gratefully acknowledge support from the CAS Center for Excellence in Particle Physics (CCEPP).

\bibliographystyle{unsrt}
\bibliography{references}

\end{document}